\documentclass[lettersize,journal]{IEEEtran}
\usepackage{amsmath,amsfonts}
\usepackage{algorithmic}
\usepackage{algorithm}
\usepackage{array}
\usepackage[caption=false,font=normalsize,labelfont=sf,textfont=sf]{subfig}
\usepackage{textcomp}
\usepackage{stfloats}
\usepackage{url}
\usepackage{verbatim}
\usepackage{graphicx}
\usepackage{cite}
\usepackage{pifont}
\usepackage{pgfplots}
\usepackage{booktabs}
\usepackage{multirow}
\usepackage{tikz}
\usetikzlibrary{positioning, shapes.geometric, arrows.meta}
\usepackage[table]{xcolor} % Required for \cellcolor
\begin{document}

\title{End-to-End Joint ASR and Speaker Role Diarization with Child-Adult Interactions}

\author{Anfeng Xu,~\IEEEmembership{Graduate Student Member,~IEEE}, Tiantian Feng, ~\IEEEmembership{Member,~IEEE}, Somer Bishop, Catherine Lord, Shrikanth Narayanan, ~\IEEEmembership{Fellow,~IEEE}
\thanks{Anfeng Xu, Tiantian Feng, and Shrikanth Narayanan are affiliated with Viterbi School of Engineering, University of Southern California, US.} \thanks{Somer Bishop is affiliated with Weill Institute for Neurosciences, University of California, San Francisco, US.} \thanks{Catherine Lord is affiliated with David Geffen School of Medicine, University of California, Los Angeles, US.}}

\markboth{Journal of \LaTeX\ Class Files,~Vol.~14, No.~8, August~2021}%
{Shell \MakeLowercase{\textit{et al.}}: A Sample Article Using IEEEtran.cls for IEEE Journals}


\maketitle

\begin{abstract}
Accurate transcription and speaker role diarization of child–adult spoken interactions are crucial for developmental and clinical research. However, manual annotation is time-consuming and challenging to scale. Existing automated systems typically rely on cascaded speaker diarization and automatic speech recognition pipelines, which can lead to error propagation. This paper presents a unified end-to-end framework that extends the Whisper encoder–decoder architecture to jointly model ASR and child–adult speaker role diarization. The proposed approach integrates: (i) a serialized output training scheme that emits speaker tags and start/end timestamps, (ii) a lightweight frame-level diarization head that enhances speaker-discriminative encoder representations, (iii) diarization-guided silence suppression for improved temporal precision, and (iv) a state-machine-based forced decoding procedure that guarantees structurally valid outputs.
Comprehensive evaluations on two datasets demonstrate consistent and substantial improvements over two cascaded baselines, achieving lower multi-talker word error rates and demonstrating competitive diarization accuracy across both Whisper-small and Whisper-large models. These findings highlight the effectiveness and practical utility of the proposed joint modeling framework for generating reliable, speaker-attributed transcripts of child–adult interactions at scale. The code and model weights are publicly available\footnote{https://github.com/usc-sail/joint-asr-diarization-child-adult}.
\end{abstract}


\section{Introduction}
Child-adult spoken interactions play a central role in developmental research, clinical assessment, and behavioral analysis. These conversational exchanges provide critical information about children’s expressive language abilities, social communication patterns, and turn-taking behavior \cite{barokova2020commentary}. 
% The behaviors of children, particularly those with neurodevelopmental disorders or delayed development, are frequently examined in interview-based assessments, such as the Autism Diagnostic Observation Schedule (ADOS) \cite{lord2000autism} and the Brief Observation of Social Communication Change (BOSSC) \cite{grzadzinski2016measuring}. 
Unlike monologic speech settings where only a single speaker’s content may be of interest, meaningful interpretation of child–adult interactions fundamentally depends on knowing what both participants say and how they respond to one another.
Extracting reliable speaker-attributed transcripts from these interactions enables the computation of conversational metrics (e.g., words per minute, utterance duration, conversational latency) that serve as quantitative indicators of spoken language development and severity of social communication symptoms  \cite{xu2023understanding, tager2009defining}. However, manual transcription and speaker labeling are extremely labor-intensive and difficult to scale for large cohorts. This motivates the development of frameworks for joint automatic speech recognition (ASR) and speaker role diarization, a variant of speaker diarization that assigns role labels (e.g., child and adult) rather than generic speaker identities, enabling accurate transcript generation with precise speaker role attribution and timing.

Despite substantial advancements in end-to-end ASR \cite{prabhavalkar2023end} and speaker diarization \cite{park2022review} systems, automated transcription of child-adult dyadic interactions remains challenging. Child speech differs significantly from adult speech in terms of pitch, formant structure, articulation, and linguistic complexity \cite{lee1999acoustics, potamianos2004robust, lee2014developmental}. These challenges are further exacerbated in naturalistic settings where recordings often contain significant ambient noise and speakers present widely varying expressive abilities \cite{li2025automated}. While such acoustic and linguistic heterogeneity poses fundamental modeling challenges for ASR systems, recent studies show that Whisper models can achieve strong performance when appropriately fine-tuned, likely due to their large-scale and diverse pretraining data~\cite{fan2024benchmarking, xu2025large}. 

Crucially, many downstream behavioral analyses require accurate utterance-level timestamps, in addition to transcripts and speaker labels \cite{shen2025conversational}. 
Prior work has approached speaker- and, in some cases, timestamp-attributed transcription using cascaded pipelines, typically performing speaker diarization followed by ASR, or applying ASR first and attributing speakers afterward \cite{sun2025said, long2024babyview, kim2025hybrid}. Diarization-first pipelines can produce reasonable speaker boundaries, but segmentation errors naturally propagate to the ASR stage. Conversely, ASR-first pipelines can degrade the performance of forced alignment and speaker attribution through transcription errors. These cascaded designs require substantial domain tuning and often produce suboptimal timestamp or speaker-role accuracy.

To overcome these challenges, we propose a \emph{unified end-to-end framework} for joint ASR and speaker-role diarization tailored to child-adult interactions. Our approach extends the Whisper encoder-decoder architecture with two key components: (1) a serialized output training scheme that generates speaker tags and start/end timestamp tokens within a single output sequence, and (2) a lightweight diarization head attached to the final encoder layer that produces frame-level speaker activity labels. Timestamp supervision, speaker-role supervision, and lexical prediction are learned jointly, enabling the encoder to develop speaker-discriminative and temporally aligned representations. To ensure structurally valid outputs, we introduce a \emph{state-machine-based forced decoding} mechanism that enforces the correct ordering of speaker and timestamp tokens. Additionally, we incorporate \emph{diarization-guided silence suppression} during decoding to reduce timestamp drift and improve boundary precision.

Our work makes the following contributions:
\begin{itemize}
    \item Unified Whisper ASR and diarization modeling: A single Whisper-based system that predicts transcripts, speaker roles, and utterance-level timestamps without external segmentation or forced alignment.
    \item Diarization-guided supervision and decoding: A frame-level diarization head shapes speaker-discriminative encoder representations, while silence-aware timestamp suppression stabilizes temporal boundaries and improves speaker-attribution consistency.
    \item Reliable structured decoding: A forced decoding state machine that eliminates missing special token errors observed in serialized output generation.
    \item Comprehensive evaluation on child--adult speech: Experiments on two child-adult interaction datasets demonstrate consistent improvements in mtWER, WER, AER, and DER over two cascaded baselines.
\end{itemize}

Overall, this work introduces a practical and effective solution for generating rich, speaker-attributed transcripts with accurate temporal boundaries in child-adult conversation settings. By integrating ASR, speaker role identification, and timestamp prediction into a single model, our approach enables scalable and reliable automatic extraction of behavioral speech measures for research and clinical applications.

\section{Related Works}
\subsection{Speaker Role Diarization for Child-Adult Interactions}
Early speaker diarization systems rely on modular pipelines that combine speech activity detection and speaker embedding extraction, such as x-vectors \cite{snyder2018x}, followed by clustering \cite{park2022review}. More recent end-to-end neural speaker diarization (EEND) methods \cite{fujita2019end, horiguchi2020end} directly predict frame-level speaker activities, achieving strong results in meeting transcription and multi-party conversations. However, little work has explored role-specific diarization in settings such as interviewer–interviewee or doctor–patient interactions. In child–adult conversations, the two speakers often exhibit highly distinctive vocal characteristics (e.g., pitch, formants, spectral profiles), making the role labels acoustically well separated \cite{lee1999acoustics, potamianos2004robust, lee2014developmental}. These innate differences render end-to-end models particularly suitable for direct speaker-role assignment, as they can learn stable, discriminative cues without relying on semantic contexts.

Building on this intuition, recent work has begun to directly target child–adult role diarization by leveraging large speech foundation models within end-to-end frameworks that predict silence, child, and adult speaker labels frame by frame \cite{xu2024exploring, li2023towards}. The studies found that pretrained models, such as Whisper \cite{radford2023robust}, WavLM \cite{chen2022wavlm}, and MMS \cite{pratap2024scaling}, substantially enhance diarization accuracy compared to traditional clustering-based systems such as PyAnnote~\cite{plaquet2023powerset} and VBx~\cite{landini2022bayesian} on clinical datasets. Notably, Whisper models consistently outperform other evaluated speech foundation models \cite{xu2024exploring}. Further extending this direction, follow-up work \cite{xu2025data} proposed a data-efficient approach that trains Whisper encoders on simulated child–adult conversations, achieving strong zero-shot performance and further gains after lightweight fine-tuning on limited real data. Together, these advances highlight both the effectiveness of large pretrained speech models and the importance of role-specific modeling tailored to dyadic child–adult interactions.

\subsection{Multi-Speaker ASR and Serialized Output Training}
Traditional speaker-attributed ASR systems are commonly built as cascaded pipelines in which diarization, separation, and ASR are handled by separate, independently trained modules \cite{watanabe2020chime, raj2021integration}. Similarly, in the child–adult domain, existing systems predominantly follow a modular approach, performing ASR child-adult speaker prediction independently \cite{sun2025said}. These pipelines suffer from error propagation between stages and incur additional computational costs due to the need for separate training for each component. 

Earlier work began exploring the idea of emitting speaker tokens jointly with lexical transcriptions \cite{shafey2019joint}. Building on this direction, Serialized Output Training (SOT) \cite{kanda2020serialized} has emerged as a prominent framework for unified multi-speaker ASR. Originally developed for end-to-end overlapped speech recognition, SOT represents multi-talker interactions as a single serialized sequence containing both lexical tokens and explicit speaker or speaker-change markers. Leveraging multi-headed attention mechanisms \cite{vaswani2017attention}, SOT-based architectures can attend to different speakers within one model rather than duplicating encoders or decoder heads. This formulation enables encoder–decoder models to jointly learn content and speaker attribution without relying on independent modular systems. Subsequent studies have extended SOT for multi-talker ASR objectives \cite{kanda2022streaming, liang2023ba, fan2024sa}, demonstrating its effectiveness in modeling multi-talker conversations within a unified sequence-to-sequence framework.

\subsection{Joint ASR and Diarization Modeling}

Following the development of the SOT framework, several works have investigated training for both ASR and speaker diarization objectives. Transcribe-to-Diarize \cite{kanda2022transcribe} builds on speaker-attributed ASR trained with SOT by learning token-level start and end times from decoder–encoder attention, enabling joint transcription and diarization, though its timestamp estimates rely on attention patterns without any supervision and are therefore less robust. Sortformer \cite{park2024sortformer} and JEDIS-LLM \cite{shi2025train} integrate speaker supervision into the ASR encoder, improving multi-speaker recognition with standard cross-entropy training, but they do not provide frame- or word-level diarization timestamps. Notably, our work is motivated by SLIDAR \cite{cornell2024one} and SOMSRED \cite{makishima2024somsred}, which jointly predict transcripts, timestamp tokens, and local speaker embeddings within a single decoding pass, followed by clustering to obtain global speaker identities. Our setting differs in that we target speaker role diarization rather than general speaker diarization, and we work in far more low-resource conditions, which necessitate fine-tuning an existing speech foundation model (e.g., Whisper) to achieve a reliable performance. In addition, following the works on LLMs~\cite{willard2023efficient}, we enforce the intended format by guiding the decoder with a state machine.

\section{Cascaded Baselines}

 We characterize the performance of non-end-to-end frameworks by evaluating cascaded systems that follow the more traditional pipeline for speaker-attributed ASR. Prior approaches typically decompose the task into modular stages, allowing speaker diarization and lexical transcription to be handled by separate systems. In this framework, we first consider a zero-shot baseline using WhisperX~\cite{bain2022whisperx}. In addition, we evaluate two fine-tuned baselines, considering using either (i) a diarization model that first predicts who is speaking and when, followed by ASR on each segmented region, or (ii) an ASR model that directly outputs both words and speaker-role tags, with segmented boundaries generated afterward via forced alignment. These two variants serve as complementary cascaded baselines against which we compare our unified end-to-end model. We choose Whisper models for both the cascaded baselines and our proposed method because they demonstrate state-of-the-art performance in both ASR and speaker role diarization in child-adult contexts, while also being suitable for SOT-based modelings \cite{xu2024exploring, xu2025large}.

\subsection{Zero-Shot: Using WhisperX}
WhisperX~\cite{bain2022whisperx} is an enhanced transcription pipeline that integrates Voice Activity Detection (VAD), Whisper ASR, and a forced alignment module based on wav2vec 2.0~\cite{baevski2020wav2vec} to generate accurate word-level timestamps. In our setup, we use WhisperX solely for transcription and timestamp prediction. Given an input audio segment, WhisperX first applies VAD to identify speech regions and then produces an utterance-level transcript using the Whisper model. Its alignment module subsequently refines these outputs by estimating start and end times for each word. This yields a sequence of (word, start time, end time) tuples, which we use as the foundation for downstream speaker-role attribution. %For the VAD, we choose to use the Silero VAD~\cite{Silero-VAD}.

To assign a speaker label to each word, we apply a publicly available child–adult diarization model~\cite{xu2025data} that outputs frame-level probabilities for child, adult, silence, and overlap classes. For each word, we identify all frames corresponding to its timestamp interval and compute the average child and adult log probabilities over the duration. The speaker role with the higher average log probability is then assigned to the word. Then, we merge words within 0.3s apart from the same speaker as a single utterance. This procedure generates utterance-level transcripts with speaker and timestamp labels, eliminating the need for additional fine-tuning, forming a strong zero-shot cascaded baseline.

\subsection{Diarization-First: Speaker Role Diarization $\rightarrow$ ASR}

\label{sec:diarization-ferst}

\begin{figure}[!t]
\centering
\includegraphics[width=\linewidth]{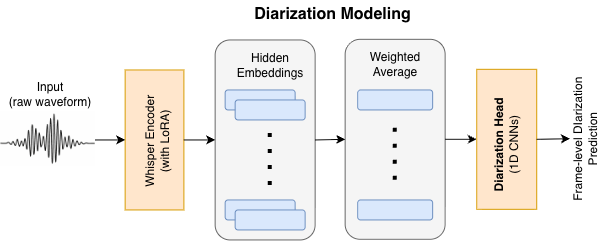}
\caption{Baseline speaker role diarization pipeline with Whisper Encoder.}
\label{fig:diarization}
\end{figure}

\label{baseline:1}
For the first fine-tuned baseline, we implement a cascaded system in which a dedicated diarization model first predicts speech segments and speaker roles. Each resulting segment is then passed to a Whisper ASR model to generate the corresponding transcript.
Frame-level speaker activities from the diarization model are first converted into contiguous segments by merging consecutive frames with the same speaker label. To produce more natural ASR input segments, predicted speech regions from the same speaker that are separated by less than 0.3s are further merged. Then, extremely short speaker segments (under 0.2s) are discarded.
Each resulting segment is then decoded independently by Whisper ASR to generate the transcript for that speaker’s turn. The ASR and diarization modules are fine-tuned independently but share the same Whisper backbone for initialization. The ASR model is trained on utterance-level inputs, while the diarization model is fine-tuned on longer speech regions, up to 30s.

The speaker role diarization modeling is illustrated in Figure~\ref{fig:diarization}, following the prior work \cite{xu2025data}. We model child-adult speaker role diarization as a frame-level classification task using the Whisper encoder.
Given an input waveform $X = [x_1, \ldots, x_T]$, the objective is to assign a label $y_t \in \{s, c, a\}$ to each frame $x_t$, corresponding to silence/noise, child speech, and adult speech, respectively.
Similar to \cite{xu2025data}, rather than unfreezing the Whisper encoder, we keep it frozen and apply Low-Rank Adaptation (LoRA)~\cite{hu2022lora} within its feed-forward transformer blocks to mitigate overfitting when fine-tuned only for the speaker role diarization objective. The input waveform is first encoded into hidden embeddings from the Whisper encoder. A learnable weighted pooling then combines the hidden embeddings from all encoder layers into a single fused embedding.
The pooled features are then fed into a stack of three 1D CNN layers, each with 256 channels, a kernel size of 1, ReLU activation, and a dropout rate of 0.1 during training.
A final 1D CNN with a kernel size of 1 and a channel width of 3 produces the frame-level class probabilities.

\subsection{ASR-First: SOT-ASR $\rightarrow$ Forced Alignment}

The second fine-tuned cascaded baseline seeks to obtain speaker-attributed transcripts without training a dedicated diarization model. We fine-tune a Whisper ASR model with SOT to generate a unified multi-speaker transcript in the form:
\[
\texttt{<speaker> transcript},
\]
where \texttt{<speaker>} can be \texttt{<child>} or \texttt{<adult>}, which mark child and adult speech, respectively. The model focuses solely on lexical and speaker-role prediction without emitting timestamp tokens. The ASR model is trained and applied to audio chunks up to 30 seconds long.
To recover temporal boundaries, we apply forced alignment between each audio chunk and its decoded transcript. Specifically, we use the NVIDIA NeMo forced aligner \cite{rastorgueva2023nemo} on the transcript after removing the speaker tokens, which yields word-level timestamps. The earliest and latest aligned word boundaries are then taken as the utterance-level start and end times.

\section{Proposed Method}
\begin{figure*}[!t]
\centering
\includegraphics[width=\textwidth]{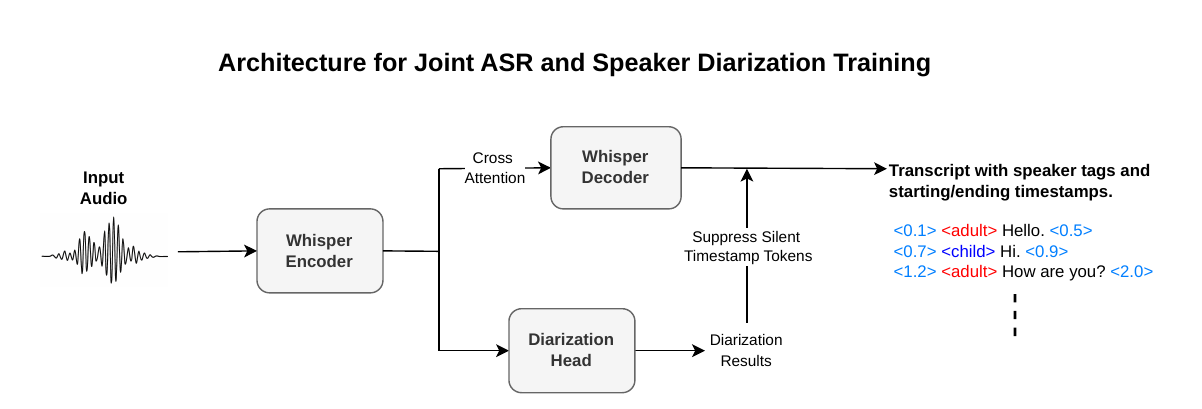}%
\label{fig_sim}
\caption{Proposed joint ASR and speaker role diarization architecture. The last hidden layer from the encoder is used for the decoder and diarization head.}
\end{figure*}

\subsection{Model Architecture}
Our framework extends the Whisper encoder-decoder architecture with both a diarization head and a serialized output training. The encoder processes input log-Mel features through a convolutional layer and a stack of transformer layers, producing contextualized embeddings that capture long-range temporal dependencies. These encoder representations are used for two complementary objectives: ASR and speaker role diarization through frame-level speaker classification.

\subsubsection{Serialized Output Training Setup}
We adopt a serialized output training (SOT) framework to handle multi-speaker conversation transcription and speaker role diarization within a single decoding stream. Specifically, we train the decoder to autoregressively produce a unified token sequence for the full conversation, including explicit speaker and timestamp tokens. Each utterance segment is represented as:
\begin{equation*}
    \texttt{<|t\_start|>} \; 
    \texttt{<speaker>} \; \textit{transcript} \; \texttt{<|t\_end|>},
\end{equation*}
where \texttt{<speaker>} indicates the active speaker (\texttt{<child>} for child, \texttt{<adult>} for adult), and \texttt{<|t\_start|>} / \texttt{<|t\_end|>} denote the starting and ending timestamps for the utterance.  
This ordering places temporal boundaries first, followed by the speaker role and then the lexical content, providing the decoder with a clear sequential structure that aligns with how utterances unfold in the audio. %Our initial experiments have shown that it yields more stable training than alternative token orderings (e.g., predicting speaker first). 
The sequence-based representation enables the model to jointly predict lexical, timestamp, and speaker-role information within a single decoding process, eliminating the need for external alignment or cascaded systems. 

\subsubsection{Diarization Head.}
In addition to the serialized output, we incorporate a lightweight diarization head attached to the final encoder layer. The head consists of a stack of 1D convolutional layers that outputs frame-level probabilities across three mutually exclusive classes: \textit{child}, \textit{adult}, and \textit{silence}, similar to the Diarization-First Baseline diarization method as detailed in~\ref{baseline:1}. This auxiliary supervision provides explicit information about speaker-role and speech activity boundaries, enhancing the encoder's speaker and acoustic representations to aid the decoder with speaker and timestamp token predictions.  
The shared encoder thus learns representations that are jointly informative for frame-level diarization and token-level ASR generation.

\subsection{Loss Function}
\label{sec:joint_training}
The model is optimized using a weighted multi-task objective that combines ASR and diarization losses:
\begin{equation}
    \mathcal{L}_{\text{total}}
    = \mathcal{L}_{\text{ASR}}
    + \mathcal{L}_{\text{diar}}.
\end{equation}

\paragraph*{ASR Loss.}
The ASR objective is the standard sequence cross-entropy loss over serialized token sequences:
\begin{equation}
    \mathcal{L}_{\text{ASR}}
    = - \frac{1}{T} \sum_{t=1}^{T}
      \log P(y_t \,|\, y_{<t}, \mathbf{h}),
\end{equation}
where $y_t$ denotes the ground-truth token (including text, timestamp, or speaker tokens) at step $t$ from $T$ total tokens, and $\mathbf{h}$ represents the encoder output.  
This formulation enables the model to learn consistent decoding of both lexical and structural tokens, aligning the ASR process with diarization and timing cues.

\paragraph*{Diarization Loss.}
The diarization head outputs a 3-dimensional probability vector
$\hat{\mathbf{s}}_n = [\hat{s}_{n}^{(\text{child})}, \hat{s}_{n}^{(\text{adult})}, \hat{s}_{n}^{(\text{sil})}]$
for each encoder frame $n$.  
We supervise these predictions using the multi-class cross-entropy loss:
\begin{equation}
    \mathcal{L}_{\text{diar}}
    = - \frac{1}{N} \sum_{n=1}^{N}
      \sum_{c \in \{\text{child},\,\text{adult},\,\text{sil}\}}
      % s_{n}^{(c)} \log \hat{s}_{n}^{(c)},
      s_{n}^{(c)} \log P(\hat{s}_{n}^{(c)} \,|\,  \mathbf{h_n}),
\end{equation}
where $s_{n}^{(c)}$ is a one-hot label for the active class at frame $n$.  
This term explicitly aligns encoder features with speaker roles and speaking activity states, promoting stronger temporal and role-specific representations that benefit both decoding and downstream diarization metrics.

\subsection{Diarization-guided Silence Suppression}
To further leverage speaker role diarization information at inference time, we introduce decoder guidance based on the diarization head outputs. 
Timestamp tokens corresponding to silence regions predicted from the diarization head are assigned 0 probabilities during the decoding process. Specifically, we identify silence regions as contiguous spans where $\hat{s}_{n}^{(\text{sil})} > 0.7$. To allow more flexible timestamp placement near boundaries, we shrink each predicted silence segment by trimming 0.2s from both its beginning and end before applying suppression.
This guidance discourages the decoder from emitting redundant timestamp tokens for silence segments, thereby reducing utterance boundary mistakes and improving utterance segmentation reliability.

\subsection{State-Machine-Based Forced Decoding}

% \begin{figure}[!t]
% \centering
% \includegraphics[width=\linewidth]{figures/sot_decode.pdf}
% \caption{State-machine diagram for forced decoding during inference.}
% \label{fig:forced_decode}
% \end{figure}

\begin{figure}[!t]
    \centering
    \scriptsize
    % Title as a separate bold text line above the diagram
    \textbf{\small State Machine for Forced Serialized Output Decoding}
    % \vspace{0.5cm} % Space between title and diagram

    \begin{tikzpicture}[
        node distance=1.2cm and 0.6cm, % vertical and horizontal spacing
        box/.style={
            rectangle,
            draw=black,
            thick,
            align=center, % Centers text and allows multiline
            minimum width=1.4cm,
            minimum height=1.0cm,
            font=\scriptsize
        },
        arrow/.style={
            ->,
            >=Latex, % Nice looking arrowhead
            thick
        }
    ]
        % --- Placement of Nodes ---
        % Place S2 first
        \node[box] (s2) {\textbf{S2} \\ Start Time};
        
        % Place S3, S4, S5 to the right of S2
        \node[box, right=of s2] (s3) {\textbf{S3} \\ Speaker};
        \node[box, right=of s3] (s4) {\textbf{S4} \\ Text};
        \node[box, right=of s4] (s5) {\textbf{S5} \\ End Time};

        % Place S1 below S2
        \node[box, below=of s2] (s1) {\textbf{S1} \\ Transcription \\ Start};
        
        % Place S6 below S5
        \node[box, below=of s5] (s6) {\textbf{S6} \\ Transcription \\ End};

        % --- Drawing Edges ---
        % Straight arrows
        \draw[arrow] (s1) -- (s2);
        \draw[arrow] (s2) -- (s3);
        \draw[arrow] (s3) -- (s4);
        \draw[arrow] (s4) -- (s5);
        \draw[arrow] (s5) -- (s6);

        % Self-loop on S4 (below the box)
        \draw[arrow] (s4) edge[loop below, looseness=4] (s4);

        % Curved feedback arrow from S5 back to S2 (above the boxes)
        % 'bend right' curves the path to the right relative to the direction of travel (S5 to S2), resulting in an upward arch.
        \draw[arrow] (s5) edge[bend right=50] (s2);

    \end{tikzpicture}
    \caption{State-machine diagram for forced decoding during inference.}
    \label{fig:forced_decode}
\end{figure}
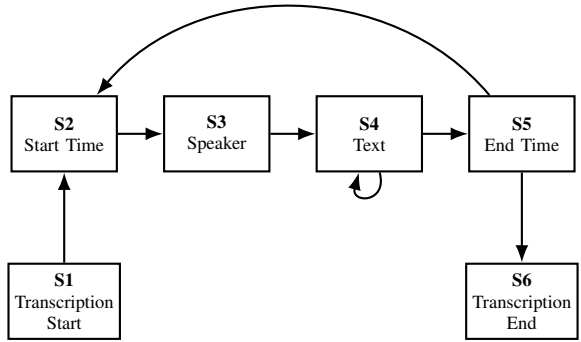

To ensure consistent generation of structured, serialized outputs, we employ a forced-decoding strategy based on a finite-state machine.  
While standard beam search decoding allows flexible token ordering, unconstrained generation may lead to malformed or misordered tokens, such as missing timestamp boundaries or incorrect speaker–text order.  
To mitigate this, we define a decoding state machine that enforces the expected token order within each conversational segment.

The state machine, as illustrated in Fig.~\ref{fig:forced_decode}, defines a valid transition graph. At each state, the decoder is expected to output the following tokens:
\begin{align*}
    S_1 &: <|startoftranscript|><|en|><|transcribe|> \\
    S_2 &: <|timestamp|> \\
    S_3 &: <child> \text{or} <adult> \\
    S_4 &: text \text{ } token \\
    S_5 &: <|timestamp|> \\
    S_6 &: <|endoftranscribe|>.
\end{align*}
During beam search, only tokens allowed by the current state are assigned nonzero probabilities, masking out invalid transitions.  
For example, once a timestamp token $<|timestamp|>$ is emitted at state $S_2$, only text tokens can follow immediately. Then, both text and ending timestamp tokens become permissible. Similarly, the next valid token after an ending timestamp must be either another speaker token or the final transcription end token.

The forced decoding framework ensures the syntactic correctness of the serialized output, guaranteeing that every speaker segment includes both start and end timestamps. When combined with the diarization-guided silence suppression described earlier, we expect to achieve more temporally and structurally coherent outputs without sacrificing speech recognition accuracy.

\subsection{Diarization Head Initialization and Fine-tuning} 

Prior to joint optimization, we initialize the diarization head through a separate pretraining stage using frozen Whisper encoder representations. We consider two input strategies during this stage: (1) the final encoder layer only (LL), and (2) a learnable weighted average of all encoder layers (WA). The first strategy is naturally aligned with the subsequent joint training stage, since the diarization head receives the final encoder-layer representations during the joint optimization. The second strategy is more closely aligned with the NNED framework as discussed in Section~\ref{sec:diarization-ferst}, where speaker-discriminative information can be aggregated across network depth rather than relying only on the top layer. It also provides an inductive bias that may encourage the encoder, during later joint fine-tuning, to preserve or elevate useful speaker-related representations into the final layer. Earlier layers often retain stronger acoustic and speaker cues, while deeper layers emphasize higher-level phonetic content, making this cross-layer combination potentially beneficial for initialization. During this pretraining stage, the encoder remains frozen, and no LoRA adaptation is applied.

After pretraining, we fine-tune the joint ASR-diarization model (Section~\ref{sec:joint_training}), using only the final encoder layer as input. This enables the encoder to adapt its top-layer features toward both token generation and speaker-role discrimination while benefiting from the pretrained initialization.
After joint training, the speaker head is further fine-tuned while all other parameters remain frozen. This step improves the accuracy of silence prediction, which is crucial for suppressing silent timestamp tokens during decoding.

\section{Experiments}
\subsection{Datasets}

\begin{table}[t]
  \caption{Summary of Playlogue and ADOS datasets.}
  \label{tab:dataset}
  \centering
  \begin{tabular*}{\linewidth}{l c c}
    \toprule
    \textbf{Category} & {\textbf{Playlogue}} & \textbf{ADOS} \\
    \cmidrule(lr){1-1}\cmidrule(lr){2-2} \cmidrule(lr){3-3}
    Age (Min-Max, in months) & 36-66 & 43-158 \\
    Gender (M/F/Unknown) & 96/61/1 & 126/45/9 \\
    Clinical Diagnosis & TD & Mostly Non-TD \\
    \cmidrule(lr){1-1} \cmidrule(lr){2-2} \cmidrule(lr){3-3}
    Segment Count (train/dev/test) & 2326/722/987 & 836/267/1082\\
    Total Hours (train/dev/test) & 16.20/5.07/6.75 & 5.36/1.66/7.05 \\
    \cmidrule(lr){1-1} \cmidrule(lr){2-2} \cmidrule(lr){3-3}
    Recording Condition & Naturalistic & Interview \\
    \cmidrule(lr){1-1} \cmidrule(lr){2-2} \cmidrule(lr){3-3}
    Publically Available & Yes & No \\
    
    \bottomrule
  \end{tabular*}
\end{table}

We consider two datasets: the publicly available Playlogue dataset \cite{kalanadhabhatta2024playlogue} and the in-house ADOS-Mod3 corpus of the Autism Diagnostic Observation Schedule \cite{bishop2016subdimensions}. Both datasets include recordings of two speaker interactions between a child and an adult. Our research adheres to all Institutional Review Board (IRB) protocols and complies with the Data Use Agreements (DUAs) by the ADOS-Mod3 data providers. The summary of the dataset statistics is summarized in Table~\ref{tab:dataset}.

\subsubsection{Playlogue}
The Playlogue dataset comprises over 33 hours of naturalistic, long-form adult–child interactions sourced from the TalkBank system. It spans three play-based corpora and one narrative corpus, all of which involve typically developed (TD) preschool-aged children (3–5 years old). These recordings capture spontaneous, free-play dyadic conversations between children and adults, including parents, clinicians, and researchers. Each recording includes word-aligned transcripts with speaker labels derived using NVIDIA NeMo forced alignment \cite{rastorgueva2023nemo}. We have verified that the corpus does not contain overlapping speech, likely because the forced aligner only supports non-overlapping segments. Since we have observed that words with abnormally long predicted durations were typically misaligned, we discard any word whose duration exceeds 2 seconds from both the audio and transcript. We merge consecutive words from the same speaker into a single utterance whenever the gaps between them are shorter than 0.3 seconds. Playlogue presents a challenging benchmark due to its real-world recording conditions, including higher background noise and less accurate ASR annotations than our in-house ADOS-Mod3 dataset. 

\subsubsection{ADOS-Mod3}
The ADOS \cite{lord2000autism} consists of multiple semi-structured activities and typically takes 40–60 minutes to administer. Module 3 (Mod3) is intended for verbally fluent children and includes 14 activities, several of which involve interview-style questions on various topics. For our analyses, we focus on two of these interview components, \textit{Social Difficulties and Annoyance} and \textit{Emotions}, totaling 352 components from 180 children. The children are generally older than those in the Playlogue dataset, with an average age of 8.53 years and a range of 3 to 13 years. Approximately half received a diagnosis of ASD, while most of the remaining participants were diagnosed with other neurodevelopmental or psychiatric conditions, including ADHD and mental or language disorders. Recordings were obtained from two medical centers: 96 from Cincinnati Children’s Hospital (CHIC) and 84 from the University of Michigan Autism and Communication Disorders Center (MICH). We also note that the majority of CHIC cases were not diagnosed with ASD, whereas the majority of MICH cases were diagnosed with ASD.

While the initial annotation provides utterance-level speaker and timestamp labels, we further refine the timestamps for speaker role diarization using NVIDIA NeMo forced alignment at the utterance level. Specifically, we adjust the boundaries to exclude leading and trailing silences and pauses longer than 0.3s, without modifying the underlying audio. Since the annotations for overlapping speech were highly unreliable and cannot be handled by the forced alignment, and to maintain consistency with the Playlogue dataset, we only consider non-overlapping speech segments in both training and testing. We refer to this dataset by ADOS for the rest of the paper for conciseness. Overall, the ADOS dataset represents a substantially different data domain from Playlogue, particularly in terms of participant age, clinical conditions, and recording setup.

\subsubsection{Data Preprocessing}
We preprocess the recordings for both datasets by placing segment boundaries at the midpoint of pauses between consecutive utterances, and greedily selecting each boundary such that the resulting segment is the longest without exceeding 30 seconds, the maximum length supported by Whisper.
As a result, the Playlogue and ADOS datasets have average segment durations of 25.0 and 23.2 seconds, respectively. We use the official train-dev-test split for the Playlogue dataset. In total, Playlogue has 16.20 hours, 5.07 hours, and 6.75 hours of training, development, and test data, respectively. For the ADOS dataset, the CHIC recordings are used exclusively for testing. The MICH recordings are partitioned at the session level into an 80:20 train/dev split. This partitioning enforces strict speaker disjointness, ensuring that no speaker in the train or development sets appears in the test set. In total, ADOS is much smaller with 5.36 hours, 1.66 hours, and 7.05 hours of training, development, and test data, respectively. 

\subsection{Metrics}

% \begin{figure}[t]
%   \centering
%   \includegraphics[width=1\linewidth]{figures/errors.png}
%   \caption{Error components for mtWER. The yellow and blue colors highlight two separate speaker roles. REF is the ground-truth transcript, while HYP is the inferred hypothesis.}
%   \label{figures:mtWER}
% \end{figure}

\begin{figure}[t]
    \centering
    \scriptsize
    \sffamily % Switch to sans-serif font to match the image style
    \renewcommand{\arraystretch}{1.5} % Add vertical padding to cells
    \setlength{\tabcolsep}{5pt}       % Adjust horizontal padding

    \begin{tabular}{|l|c|c|c|c|c|c|c|c|}
        \hline
        REF & 
            & \cellcolor{yellow}how 
            & \cellcolor{yellow}are 
            & \cellcolor{yellow}you 
            & \cellcolor{cyan}I 
            & \cellcolor{cyan}am 
            & \cellcolor{cyan}good 
            & \cellcolor{cyan}thanks \\ 
        \hline
        HYP & \cellcolor{yellow}oh 
            & \cellcolor{yellow}how 
            & \cellcolor{yellow}were 
            & \cellcolor{cyan}you 
            & \cellcolor{cyan}I 
            & \cellcolor{cyan}am 
            & \cellcolor{yellow}great 
            & \\ 
        \hline
        ERROR & INS 
              & - 
              & SUB 
              & ATTR 
              & - 
              & - 
              & \shortstack{SUB ATTR} % Stacks the text
              & DEL \\ 
        \hline
    \end{tabular}
    
    \caption{Error components for mtWER. The yellow and blue colors highlight two separate speaker roles. REF is the ground-truth transcript, while HYP is the inferred hypothesis.}
    \label{fig:mtwer}
\end{figure}

We employ the \textbf{multi-talker Word Error Rate (mtWER)}, originally introduced in the CHiME-8 MMCSG Challenge~\cite{zmolikova2024chime}. This metric extends conventional WER by jointly evaluating transcription accuracy and speaker-role-attribution consistency in multi-speaker conditions. In addition to standard word errors (substitutions, insertions, deletions), mtWER explicitly counts speaker-role-attribution mistakes, making it a more informative metric for joint speaker-role-attributed ASR. An illustrative example of all error components is shown in Figure~\ref{fig:mtwer}. We report mtWER averaged over the child and adult speaker roles. The metric for each role is computed as
\[
mtWER_{s} = \frac{INS_{s} + DEL_{s} + SUB_{s} + ATTR_{s}}{NREF_{s}},
\]
where $s$ denotes a specific speaker role (e.g., child), and $INS_{s}, DEL_{s}, SUB_{s}$, and $ATTR_{s}$ are the insertion, deletion, substitution, and speaker-role-attribution errors, respectively. Here, $NREF_{s}$ is the total number of reference words from speaker $s$. Speaker-role-attribution errors occur when a word spoken by $s$ is incorrectly assigned to another speaker role, including cases where the word itself is substituted.

In addition to mtWER, we also report the \textbf{WER} and \textbf{Attribution Error Rate (AER)}, each computed per speaker role:
\[
WER_{s} = \frac{INS_{s} + DEL_{s} + SUB_{s}}{NREF_{s}}, 
AER_{s} = \frac{ATTR_{s}}{NREF_{s}},
\]
where $s$ denotes the specific speaker role. Similar to mtWER, we report WER and AER averaged over child and adult speaker roles.

We also report the \textbf{Diarization Error Rate (DER)}, which evaluates speaker role diarization accuracy by accounting for missed speech, false alarms, and speaker confusion errors. Formally,
\[
DER = \frac{MD + FA + SC}{TOTAL},
\]
where $MD$, $FA$, and $SC$ denote Missed Detection, False Alarm, and Speaker Confusion, respectively. We note that Playlogue~\cite{kalanadhabhatta2024playlogue} reports DER using the total audio duration (including silence) as $TOTAL$, whereas we follow the standard definition using only reference speech. We use Pyannote metrics \cite{bredin2017pyannote} to compute the DERs.

For the final results, we report the average scores for each of the above metrics across all files.

\subsection{Experimental Setup}
We use a single NVIDIA
RTX A6000 48GB GPU for all the experiments. We choose Whisper-small (English only) and Whisper-large (v3) models from HuggingFace~\cite{wolf2019huggingface} for our experiments. We perform all analyses and ablation experiments using Whisper-small to enable broad comparisons across training variants. For downstream applications, we use Whisper-large to obtain higher absolute performance. When training for ASR objectives, we apply the Whisper English text normalizer to the texts as a pre-processing step. For decoding, we use greedy decoding with a repetition penalty of 1.1. Infinite decoding loops are detected by the decoder reaching the maximum token length of 256, in which case the final utterance is discarded. We specify other experimental details for the proposed method and the baselines below.

\subsubsection{Proposed Method}
For the diarization head pretraining and fine-tuning, we train for up to 10 epochs with a batch size of 16. We use the Adam optimizer with a learning rate of $2e^{-4}$. The best model is chosen based on the validation loss, evaluated at the end of every epoch. We use a weight decay of $1e^{-5}$ during fine-tuning and a larger value of $0.01$ during pretraining to mitigate overfitting.

For the joint-training stage, we also use a batch size of 16 and the Adam optimizer, but for 20 epochs with learning rates of $5e^{-6}$ and $1e^{-6}$ for Whisper-small and Whisper-large models, respectively. All the parameters from the encoder, decoder, and diarization head are unfrozen. The model checkpoints with the lowest validation loss are selected and evaluated at the end of every epoch. Validation loss is used instead of validation WER, as computing WER is substantially more expensive.

\subsubsection{Diarization-First Baseline} For the speaker role diarization model, we apply LoRA with rank 16 to the feed-forward layers of the transformer, and use the same remaining hyperparameters as in the diarization-head fine-tuning of the proposed joint method. For the ASR models, since utterance-level sampling yields a significantly larger number of training instances, we use a batch size of 32 and evaluate every 200 steps, rather than at the end of each epoch. Training runs for up to 5 epochs, which we find sufficient for convergence. All other hyperparameters match those used during joint training in the proposed method.

\subsubsection{ASR-First Baseline} For training ASR with SOT, we use the same hyperparameters as those used for the joint training stage for the proposed method, with both the encoder and decoder unfrozen. For the forced alignment with Nvidia Nemo, we use the default hyperparameters with the \textit{stt\_en\_fastconformer\_hybrid\_large\_pc} model.

\section{Main Results}

\begin{table*}[t]
\centering
\caption{Main Results against Baselines. The results are reported in percentages, with the
lower and upper bounds of the confidence interval (CI) between 2.75\% and 97.5\% for each result (lower bound, upper bound).}
\label{tab:main-results}
\begin{tabular}{l l l c c c c c c}
\toprule
\textbf{Dataset} & \textbf{Model} & \textbf{Method} & \textbf{E2E} & \textbf{mtWER} & \textbf{WER} & \textbf{AER} & \textbf{DER}  \\
\cmidrule(lr){1-1} \cmidrule(lr){2-2} \cmidrule(lr){3-3} \cmidrule(lr){4-4} \cmidrule(lr){5-7} \cmidrule(lr){8-8}
% \midrule
\multirow{8}{*}{Playlogue} &
\multirow{4}{*}{Whisper-small} 
 & Baseline (Zero-Shot) & \ding{55} & $94.4 (83.2, 105.7)$ & $89.9 (78.7, 101.2)$ & $4.5 (3.6, 5.4)$ & $65.4 (63.4, 67.4)$ \\
 & & Baseline (Diarization-First) & \ding{55} & $45.4 (40.0, 50.7)$ & $43.5 (38.3, 48.7)$ & $\mathbf{1.9 (1.4, 2.3)}$ & $\mathbf{35.7 (33.2, 38.1)}$ \\
 & & Baseline (ASR-First) & \ding{55} & $41.4 (36.4, 46.4)$ & $38.0 (32.8, 43.1)$ & $3.4 (2.5, 4.3)$ & $56.1 (51.3, 60.9)$ \\
 & & Proposed & \ding{51} & $\mathbf{37.4 (32.7,42.1)}$ & $\mathbf{35.5 (30.8, 40.1)}$ & $\mathbf{1.9} (1.5, 2.4)$ & $40.6 (37.5, 43.8)$ \\
\cmidrule(lr){2-2} \cmidrule(lr){3-3} \cmidrule(lr){4-4} \cmidrule(lr){5-7} \cmidrule(lr){8-8}
 & \multirow{4}{*}{Whisper-large} 
 & Baseline (Zero-Shot) & \ding{55} & $91.7 (80.3, 103.1)$ & $87.0 (75.7, 98.3)$ & $4.7 (3.8, 5.6)$ & $64.4 (62.4, 66.5)$ \\
 & & Baseline (Diarization-First) & \ding{55} & $38.8 (34.3, 43.3)$ & $37.2 (32.9, 41.6)$ & $\mathbf{1.6 (1.2, 1.9)}$ & $\mathbf{36.4 (34.0, 38.9)}$ \\
 & & Baseline (ASR-First) & \ding{55} & $47.5 (42.7, 52.3)$ & $41.8 (37.1, 46.5)$ & $5.7 (4.4, 7.0)$ & $67.4 (62.9, 71.9) $ \\
 & & Proposed & \ding{51} & $\mathbf{34.3 (30.0, 38.7)}$ & $\mathbf{32.2 (28.1, 36.3)}$ & $2.1 (1.5, 2.7)$ & $42.6 (39.7, 45.6)$ \\
  \cmidrule(lr){1-1} \cmidrule(lr){2-2} \cmidrule(lr){3-3} \cmidrule(lr){4-4} \cmidrule(lr){5-7} \cmidrule(lr){8-8}
  \multirow{8}{*}{ADOS} &  
 \multirow{4}{*}{Whisper-small} 
 & Baseline (Zero-Shot) & \ding{55} & $93.9 (83.9, 104.0)$ & $85.0 (75.0, 95.0)$ & $8.9 (7.5, 10.3)$ & $74.9 (71.4, 78.4)$ \\
 & & Baseline (Diarization-First) & \ding{55} & $36.1 (33.2, 39.0)$ & $35.1 (32.2, 37.9)$ & $\mathbf{1.0} (0.8, \mathbf{1.3})$ & $\mathbf{20.4 (19.2, 21.6)}$  \\
 & & Baseline (ASR-First) & \ding{55} & $33.6 (30.1, 37.1)$ & $29.6 (26.1, 33.1)$ & $4.1 (3.2, 4.9)$ & $37.7 (36.1, 39.3) $ \\
 & & Proposed & \ding{51} & $\mathbf{28.8 (25.9, 31.6)}$ & $\mathbf{27.8 (25.0, 30.6)}$ & $\mathbf{1.0 (0.7, 1.3)}$ & $21.8 (20.4, 23.1)$ \\
\cmidrule(lr){2-2} \cmidrule(lr){3-3} \cmidrule(lr){4-4} \cmidrule(lr){5-7} \cmidrule(lr){8-8}
 & \multirow{4}{*}{Whisper-large} 
 & Baseline (Zero-Shot) & \ding{55} & $95.8 (84.6, 107.0)$ & $86.1 (75.0, 97.3)$ & $9.7 (8.2, 11.1)$ & $73.4 (70.0, 76.8)$ \\
 & & Baseline (Diarization-First) & \ding{55} & $29.7 (27.0, 32.4)$ & $28.6 (26.0, 31.3)$ & $\mathbf{1.0 (0.7, 1.3)}$ & $19.2 (18.0,20.5)$ \\
 & & Baseline (ASR-First) & \ding{55} & $27.0 (24.6, 29.5)$ & $22.4 (20.1, 24.7)$ & $4.6 (3.6, 5.6)$ & $36.7 (35.1, 38.4)$ \\
 & & Proposed & \ding{51} & $\mathbf{21.7 (19.4, 23.9)}$ & $\mathbf{20.6 (18.3, 22.8)}$ & $1.1 (\mathbf{0.7}, 1.6)$ & $\mathbf{18.4 (17.1, 19.6)}$ \\
\bottomrule
\end{tabular}
\end{table*}

\begin{table}[t]
  \caption{WER ($\%$) for Diarization-First approach with Ground-truth diarization regions. The parenthesis ($\downarrow$) shows the absolute improvements over using inferred diarization results.}
  \label{tab:gt-diarization}
  \centering
  \begin{tabular*}{0.64\linewidth}{l c c}
    \toprule
    {\textbf{Model}} & \textbf{Playlogue} & \textbf{ADOS} \\
    \cmidrule(lr){1-1}\cmidrule(lr){2-3}
    Whisper-small& $42.4(\downarrow 1.1)$ & $26.3(\downarrow 8.8)$ \\
    Whisper-large & $38.7(\uparrow 1.5)$ & $22.1(\downarrow 6.5)$ \\
                          
    \bottomrule
  \end{tabular*}
\end{table}

Table~\ref{tab:main-results} summarizes the ASR and Diarization performances of all systems across the Playlogue and ADOS datasets, along with the $95\%$ confidence intervals to show the statistical reliability and variability of the reported metrics.

\subsection{Baseline Results}
The zero-shot results show very high ASR and diarization error rates, underscoring the need for domain-specific fine-tuning. Comparing the Diarization-First and ASR-First fine-tuned baselines reveals a clear trade-off in the cascaded systems. Diarization-First Baseline segments audio using a dedicated diarization model before transcription, yielding more accurate speaker boundaries. However, its ASR performance is limited by error propagation, as segmentation mistakes degrade the subsequent transcription. In contrast, ASR-First Baseline benefits from applying ASR first, achieving lower WER in most setups. Yet its AER is noticeably higher, reflecting the difficulty of naively relying on SOT-based ASR outputs for speaker-role detection. In addition, the forced alignment step is limited by the quality of the ASR hypotheses. Transcription inaccuracies propagate into the alignment process, resulting in less reliable timestamp predictions and much higher DER. 

Table~\ref{tab:gt-diarization} reports WERs for the diarization-first baseline using oracle (ground-truth) speaker-role segments. For ADOS, where the diarization annotations are relatively clean, using oracle segments substantially improves performance, highlighting the effect of error propagation in the diarization-first pipeline. For Playlogue, however, the results are comparable to those obtained with inferred diarization, likely because the manual annotations are noisier and the diarization system can smooth some of this noise while still introducing its own errors. Notably, these WERs are comparable to those of the ASR-first baseline, suggesting that utterance- or chunk-level (up to 30 seconds) ASR yields similar performance.

\subsection{Proposed Method -- mtWER}
Overall, the proposed joint ASR and Speaker Role Diarization framework shows consistent mtWER reductions over the cascaded baselines across both Whisper-small and Whisper-large configurations, achieving approximately $9.7\sim 19.6\%$ relative reductions in mtWER. Compared to the Diarization-First Baseline, the joint approach avoids the error propagation introduced by external segmentation, where boundary mistakes directly disrupt downstream transcription. Relative to ASR-First Baseline, the model benefits from explicit frame-level speaker-role supervision, which mitigates the speaker-token errors in SOT-ASR outputs, as observed by AER. Notably, the joint model also improves WER compared to the ASR-First Baseline. We attribute this to the timestamp-prediction mechanism and joint optimization, which together produce more temporally aligned and coherent encoder representations, ultimately leading to more robust decoder performance.
Overall, jointly modeling ASR and speaker roles provides a stronger inductive bias for producing coherent speaker-labeled transcripts, thereby improving both WER and the speaker-attribution component within mtWER.

\subsection{Proposed Method -- DER}
The proposed joint method consistently outperforms ASR-First Baseline in DER, as it avoids the forced-alignment stage where transcription errors propagate into boundary predictions. However, the DER trend becomes more nuanced when compared to the Diarization-First Baseline. On the Playlogue dataset, the joint model yields higher DER than the cascaded Diarization-First Baseline for both Whisper-small and Whisper-large. This outcome is expected given the challenging recording conditions, where environmental noise and annotation inconsistencies make timestamp token decoding particularly sensitive to boundary errors, whereas the dedicated diarization model in Diarization-First Baseline remains more robust.
In contrast, on the ADOS dataset, the DER gap between the joint model and Diarization-First Baseline is substantially smaller for Whisper-small, and the joint model even surpasses Diarization-First Baseline when using Whisper-large. ADOS recordings are cleaner and more structured, reducing the ambiguity inherent in timestamp prediction. Under these conditions, the shared encoder representation in the joint framework becomes advantageous, enabling the model to match the diarization performance of a separate, specialized diarization system.

\subsection{How difficult is child ASR compared to adult?}
\begin{table}[t]
  \caption{Individual mtWER results for child and adult. The parenthesis ($\downarrow$) shows the absolute improvements over the best baseline result. Results are shown in percentage($\%$).}
  \label{tab:child-adult}
  \centering
  \begin{tabular*}{0.81\linewidth}{l c c c}
    \toprule
    \textbf{Dataset} & {\textbf{Model}} & \textbf{Child} & \textbf{Adult} \\
    \cmidrule(lr){1-1}\cmidrule(lr){2-2} \cmidrule(lr){3-4}
    \multirow{2}{*}{Playlogue} & Whisper-small& $53.1(\downarrow 4.1)$ & $21.7(\downarrow 3.9)$ \\
    & Whisper-large & $48.4(\downarrow 5.3)$ & $19.8(\downarrow 4.1)$ \\
    \cmidrule(lr){1-1} \cmidrule(lr){2-2} \cmidrule(lr){3-4} 
    \multirow{2}{*}{ADOS} & Whisper-small & $45.0(\downarrow 3.9)$ & $12.5(\downarrow 5.9)$ \\
    & Whisper-large & $35.0(\downarrow 5.2)$ & $8.4(\downarrow 5.5)$ \\
                          
    \bottomrule
  \end{tabular*}
\end{table}

\begin{table}[t]
  \caption{Child mtWER from ADOS based on the age and ADOS CSS scores. Results are shown in percentage($\%$). Small and Large indicate Whisper-small and Whisper-large, respectively.}
  \label{tab:child-demo}
  \centering
  \begin{tabular*}{0.75\linewidth}{l c c c}
    \toprule
    \multirow{2}{*}{\textbf{Demographics}} & \textbf{Session} & \multicolumn{2}{c}{\textbf{mtWER}} \\
    & \textbf{Count} & \textbf{Small} & \textbf{Large} \\
    \cmidrule(lr){1-1}\cmidrule(lr){2-2} \cmidrule(lr){3-4}
    low CSS \& $\leq 8$ years & $40$ & $43.7$ & $33.3$\\
    low CSS \& $> 8$ years & $68$ & $43.1$ & $32.8$\\
    \cmidrule(lr){1-1} \cmidrule(lr){2-2} \cmidrule(lr){3-4} 
    high CSS \& $\leq 8$ years & $22$ & $48.8$ & $42.0$\\
    high CSS \& $> 8$ years & $20$ & $49.4$ & $38.8$\\
                          
    \bottomrule
  \end{tabular*}
\end{table}

Table~\ref{tab:child-adult} presents a comparison of the mtWER between children and adults. Child speech shows markedly higher mtWER on both Playlogue and ADOS, highlighting the persistent modeling challenges posed by children’s less stable articulation patterns and the fact that Whisper is likely primarily trained on adult speech. Adult speech, by contrast, is recognized with much lower error rates. For both children and adults, Whisper-large consistently outperforms Whisper-small, indicating that the larger model is better at transcribing both child and adult speech. Overall, our method consistently improves mtWERs by approximately from $4\%$ to $6\%$ absolute reductions over the best baseline results for both child and adult speech across different models and different datasets.

Table~\ref{tab:child-demo} reports child mtWER broken down by demographic characteristics. Children are divided into four groups based on age (8 years and younger/older than 8 years) and the ADOS Calibrated Severity Score (CSS), a standardized 10-point continuous metric that reflects the severity of autism-related behaviors observed during the ADOS assessment. CSS scores of 6 or higher are typically associated with severe autism behaviors, whereas scores of 3 or lower indicate few atypical behaviors. Accordingly, we partition CSS into low (1–3) and high (6–10) severity groups. Across both age groups, children with high CSS consistently exhibit higher mtWER, whereas mtWER remains comparable within each age group at the same severity level. These results suggest that increased autism symptom severity is a primary factor contributing to recognition difficulty, reflecting more challenging speech characteristics for the model.

\subsection{Summary of Main Findings}
The joint model provides the best overall performance across transcription and speaker-role metrics. It consistently improves mtWER relative to both cascaded baselines and delivers strong diarization accuracy—substantially outperforming the ASR-First Baseline and approaching the Diarization-First Baseline, depending on the dataset conditions. These results highlight the advantage of integrating ASR and diarization within a single framework, reducing error propagation and producing more coherent, speaker-attributed transcripts.

\section{Analysis}
To understand how timestamp supervision and diarization supervision reshape the acoustic representation space, we compare three configurations in the following analyses: (1) \emph{Without Timestamp (Without T)}, (2) \emph{With Timestamp (With T)}, and (3) \emph{With Timestamp and Diarization Head (With T and DH)}. For (3), we show the results for both when the diarization head (DH) is pretrained with the last layer (LL) and the weighted average (WA) of all hidden representations. Note that the Without T configuration is identical to the ASR-SOT used for the ASR-First Baseline.

\subsection{k-NN Classification Accuracy}
\begin{figure}[!t]
\centering
\includegraphics[width=3in]{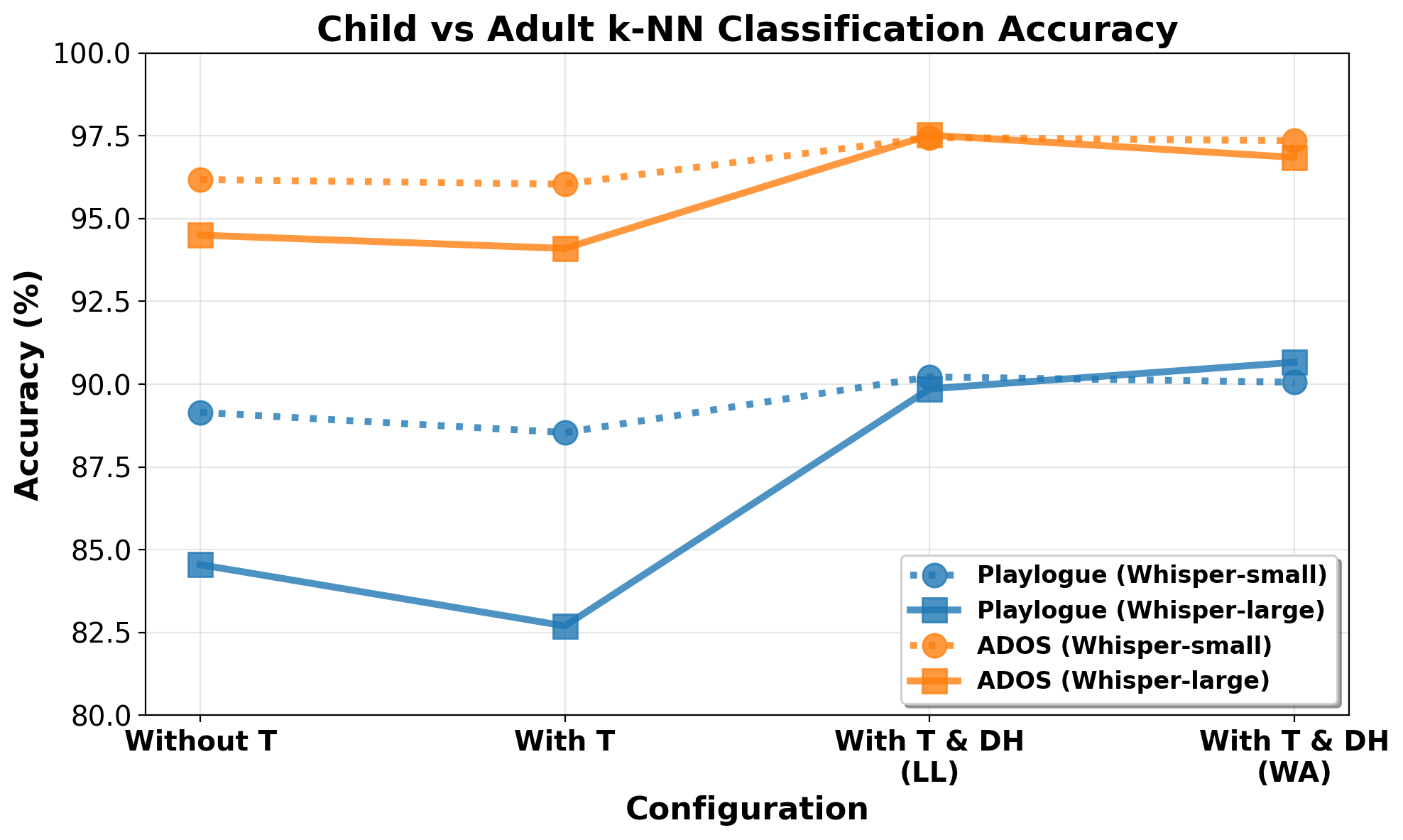}
\caption{Child vs adult kNN classification on utterance-level encoder outputs. T and DH denote Timestamp and Diarization Head, respectively. LL and WA denote Last Layer and Weighted Average during DH pretraining.}
\label{fig:knn}
\end{figure}

% \begin{figure}[t]
%     \centering
%     \begin{tikzpicture}
%         \begin{axis}[
%             title={\textbf{Child vs Adult k-NN Classification Accuracy}},
%             xlabel={\textbf{Configuration}},
%             ylabel={\textbf{Accuracy (\%)}},
%             xmin=0.8, xmax=3.2,
%             ymin=85, ymax=100,
%             xtick={1,2,3},
%             xticklabels={Without T, With T, With T and DH},
%             ymajorgrids=true,
%             grid style={dashed, gray!30},
%             legend style={at={(0.98,0.98)}, anchor=north east, draw=gray!80, font=\footnotesize},
%             width=8.5cm, % Adjusted for typical column width
%             height=6cm,
%             mark size=2.5pt,
%             label style={font=\small},
%             tick label style={font=\small}
%         ]

%         % Plot for Playlogue (Blue)
%         \addplot[
%             color=cyan!70!blue,
%             mark=*,
%             thick
%         ]
%         coordinates {
%             (1,89.1)
%             (2,88.5)
%             (3,90.0)
%         };
%         \addlegendentry{Playlogue}

%         % Plot for ADOS (Orange)
%         \addplot[
%             color=orange,
%             mark=*,
%             thick
%         ]
%         coordinates {
%             (1,96.1)
%             (2,96.0)
%             (3,97.3)
%         };
%         \addlegendentry{ADOS}

%         \end{axis}
%     \end{tikzpicture}
%     \caption{Child vs adult kNN classification on utterance-level encoder outputs. T and DH denote Timestamp and Diarization Head, respectively.}
%     \label{fig:knn_results}
% \end{figure}

To examine how different training configurations shape the encoder’s speaker-related representations, we evaluate child–adult separability using k-nearest neighbor (k-NN) classification to classify child-adult speaker roles from the encoder outputs. k-NN classification is widely used as a standard non-parametric probe of representation quality in self-supervised learning, assessing whether samples with similar semantics form locally consistent neighborhoods without introducing additional learnable parameters \cite{caron2021emerging}. This is performed at the utterance level, where speech representations are averaged over the temporal dimension for each utterance. We use all utterances from the test set for each dataset. 

Fig.~\ref{fig:knn} summarizes the k-NN accuracies for the four configurations with $k = 5$ and cosine distance.
Overall, the results reveal a consistent trend across both datasets. Introducing timestamp prediction slightly reduces k-NN accuracy in both datasets, suggesting that the added temporal supervision introduces variability in the encoder space that is not directly aligned with speaker identity.
In contrast, adding the diarization head with both pretraining objectives restores speaker-discriminative information, yielding substantially higher k-NN accuracies on both datasets and leading to a speaker-informative representation space. Notably, the Whisper-large models show worse child-adult separability without the diarization head, likely because the deeper architecture strengthens phonetic and semantic abstraction, which can suppress speaker-specific cues unless explicitly guided. Together, these results highlight a complementary relationship between timestamp-guided temporal alignment and diarization-driven speaker discrimination.

\subsection{Linear Probing for Encoder Temporal Alignment}
% \begin{table}[t]
%   \caption{Speaker classification linear probing for each 0.1s on 0.3s around the speech boundaries. T and DH denote Timestamp and Diarization Head. Results shown in percentage (\%)}
%   \label{tab:linearprobe}
%   \centering
%   \begin{tabular}{l c c c}
%     \toprule
%     \multirow{2}{*}{\textbf{Dataset}} & \multirow{2}{*}{\textbf{Config}} & \textbf{F1}  & \multirow{2}{*}{\textbf{Acc}} \\
%      &  & \textbf{Child / Adult / Silence / Macro} & \\
    
%     \cmidrule(lr){1-1} \cmidrule(lr){2-2} \cmidrule(lr){3-3} \cmidrule(lr){4-4} 
%     \multirow{3}{*}{Playlogue} & w/o T & $70.68$ / $60.97$ / $72.68$ / $68.11$ & $69.10$ \\
%      & w/ T & $70.83$ / $60.98$ / $73.19$ / $68.33$ & $69.37$ \\
%      & w/ T \& DH & $\mathbf{72.74}$ / $\mathbf{63.46}$ / $\mathbf{74.20}$ / $\mathbf{70.13}$ & $\mathbf{70.98}$ \\

%     \cmidrule(lr){1-1} \cmidrule(lr){2-2} \cmidrule(lr){3-3} \cmidrule(lr){4-4} 
      
%     \multirow{3}{*}{ADOS} & w/o T & $81.93$ / $76.36$ / $76.64$ / $79.27$ & $79.40$ \\
%      & w/ T & $82.35$ / $76.74$ / $79.60$ / $79.56$ & $79.66$ \\
%      & w/ T \& DH & $\mathbf{85.13}$ / $\mathbf{78.94}$ / $\mathbf{80.09}$ / $\mathbf{81.38}$ & $\mathbf{81.29}$ \\
%     \bottomrule
%   \end{tabular}
% \end{table}

\begin{table}[t]
  \caption{Speaker classification linear probing for each 0.1s on 0.3s around the speech boundaries. T and DH denote Timestamp and Diarization Head. Results shown in percentage (\%)}
  \label{tab:linearprobe}
  \centering
  \begin{tabular}{l c c c}
    \toprule
    \multirow{2}{*}{\textbf{Dataset}} & \multirow{2}{*}{\textbf{Config}} & \textbf{F1}  & \multirow{2}{*}{\textbf{Acc}} \\
      &  & \textbf{Child/Adult/Silence/Macro} & \\

     \cmidrule(lr){1-1} \cmidrule(lr){2-2} \cmidrule(lr){3-3} \cmidrule(lr){4-4} 
     \multirow{4}{*}{Playlogue} & w/o T & $69.6$ / $59.5$ / $72.4$ / $67.2$ & $68.3$ \\
      & w/ T & $70.0$ / $60.0$ / $72.8$ / $67.6$ & $68.7$ \\
      & w/ T \& DH (LL) & $\mathbf{71.7}$ / $\mathbf{61.8}$ / $\mathbf{73.2}$ / $\mathbf{68.9}$ & $\mathbf{69.8}$ \\
     & w/ T \& DH (WA) & $\mathbf{71.7}$ / $61.1$ / $73.1$ / $68.6$ & $69.6$ \\

    \cmidrule(lr){1-1} \cmidrule(lr){2-2} \cmidrule(lr){3-3} \cmidrule(lr){4-4}

     \multirow{4}{*}{ADOS} & w/o T & $82.2$ / $75.5$ / $78.7$ / $78.8$ & $78.9$ \\
      & w/ T & $82.2$ / $75.2$ / $78.9$ / $78.8$ & $78.9$ \\
      & w/ T \& DH (LL) & $\mathbf{84.1}$ / $\mathbf{78.1}$ / $\mathbf{79.2}$ / $\mathbf{80.5}$ & $\mathbf{81.2}$ \\
     & w/ T \& DH (WA) & $83.8$ / $77.4$ / $78.4$ / $79.9$ & $79.7$ \\
     
    \bottomrule
  \end{tabular}
\end{table}

To assess how timestamp and diarization supervision influence encoder representations, particularly their temporal alignment with the input, we perform linear probing by fitting logistic regression models on encoder embeddings near utterance boundaries. Specifically, we extract encoder outputs from 0.6s audio windows spanning from 0.3s before to 0.3s after the child and adult utterance boundaries. We compute the average embeddings over 5 frames (0.1 seconds) at each boundary from Whisper models, resulting in 6 samples per utterance boundary. The boundaries are extracted from 200 randomly sampled audio segments (up to 30 seconds) for both the training and test sets. 

As shown in Table~\ref{tab:linearprobe}, adding timestamp supervision provides small improvements in silence F1 scores and better temporal alignment of voice activity, but overall, the improvements are limited. However, incorporating the diarization head with both pretraining objectives improves all metrics across both datasets, with especially notable gains for child and adult F1 scores. These results suggest that the diarization head supervision enhances boundary-level temporal alignment in the encoder, particularly strengthening speaker-role discrimination at those same transitions.

\subsection{Performance Degradation under Overlapped Speech}
\begin{table}[t]
  \caption{Overlap error analysis under controlled overlap conditions (None, 0.2s, 0.5s, 1.0s). Metrics include mtWER, WER breakdown (INS/DEL/SUB), AER, and DER (\%).}
  \label{tab:overlap}
  \centering
  \begin{tabular}{l c c c c c}
    \toprule
    \multirow{2}{*}{\textbf{Whisper}} & \multirow{2}{*}{\textbf{Overlap}} & \multirow{2}{*}{\textbf{mtWER}} & \textbf{WER} & 
    \multirow{2}{*}{\textbf{AER}} & \multirow{2}{*}{\textbf{DER}} \\
      &  & & \textbf{INS / DEL / SUB} & \\
     \cmidrule(lr){1-1} \cmidrule(lr){2-2} \cmidrule(lr){3-5} \cmidrule(lr){6-6} 
     \multirow{4}{*}{Small} & None & $33.8$ & $7.3$ / $12.1$ / $11.2$ & $3.1$ & $22.2$ \\
      & 0.2s & $40.0$ &  $6.7$ / $18.6$ / $10.4$ & $4.5$ & $29.0$ \\
      & 0.5s & $42.8$ & $5.0$ / $23.3$ / $10.1$ & $4.3$ & $34.8$ \\
     & 1.0s & $47.2$ & $3.8$ / $28.2$ / $10.2$ & $5.1$ & $42.6$ \\
     \cmidrule(lr){1-1} \cmidrule(lr){2-2} \cmidrule(lr){3-5} \cmidrule(lr){6-6} 
     \multirow{4}{*}{Large} & None & $27.8$ & $6.7$ / $6.2$ / $9.8$ & $5.1$ & $15.5$ \\
      & 0.2s & $33.2$ &  $6.6$ / $11.0$ / $9.0$ & $6.6$ & $21.9$ \\
      & 0.5s & $36.5$ & $5.1$ / $15.3$ / $8.7$ & $7.3$ & $29.6$ \\
     & 1.0s & $46.2$ & $3.6$ / $23.3$ / $9.0$ & $10.2$ & $39.4$ \\

    \bottomrule
  \end{tabular}
\end{table}
Analyzing overlapped speech is crucial, as it remains a major challenge for ASRs in real-world conversational settings. Table~\ref{tab:overlap} summarizes the results under controlled overlapping speech conditions. We construct the evaluation set by extracting consecutive utterance pairs from the ADOS test set with different speakers, requiring each utterance to be at least 2 seconds long and the inter-utterance pause to be less than 5 seconds, yielding 505 utterance pairs. We then simulate overlap by first removing the original pause and overlaying the two utterances with fixed overlap durations of 0.2, 0.5, and 1.0 seconds, while also retaining the original (non-overlap) condition for comparison. To reduce positional bias of utterances starting exactly at the beginning of audio, we prepend a leading silence of random duration uniformly sampled between 0 and 1 seconds to each constructed sample.

The overlap error analysis reveals a clear and consistent degradation in performance as the degree of overlapping speech increases for both Whisper-small and Whisper-large models. For Whisper-small, mtWER rises substantially from 33.8\% under no-overlap conditions to 47.2\% at 1.0s overlap, with corresponding increases in DER from 22.2\% to 42.6\%. This trend is primarily driven by a sharp increase in deletion errors (from 12.1\% to 28.2\%), suggesting that the model increasingly fails to capture portions of speech when overlap becomes more severe. Similarly, Whisper-large exhibits notable degradation, where mtWER increases from 27.8\% to 46.2\%, and DER from 15.5\% to 39.4\% as overlap increases. While insertion and substitution errors remain relatively stable, deletion errors again dominate the performance drop, indicating that overlapping speech mainly leads to missed speech segments rather than incorrect or hallucinated transcriptions. Overall, these results highlight the challenges of the proposed method trained with non-overlapping speech, with deletion errors being the primary contributors. While the current study is limited by the availability of reliable overlapping speech annotations, future work can investigate overlap-aware dataset construction and training strategies.

\section{Ablation}
\subsection{Ablation on Diarization Head}
\begin{table}[t]
  \caption{Ablation on diarization head using Playlogue data. None means no diarization head is attached. Proposed uses pretrained (weighted average) with silence suppression.}
  \label{tab:dh-ablation-play}
  \centering
  \begin{tabular}{l l c c}
    \toprule
    \textbf{Whisper} & 
    \textbf{Disarization Head} & \textbf{mtWER/WER/AER} & \textbf{DER} \\
    \cmidrule(lr){1-1} \cmidrule(lr){2-4}  
     & None & $40.7$ / $37.9$ / $2.8$ & $43.3$ \\
     & Random  & $38.3^{\dagger}$ / $36.3^{\dagger}$ / $2.0^{\dagger}$ & $42.3$ \\
     Small & Pretrained (last-layer) & $39.9$ / $37.5$ / $2.3^*$ & $42.2^*$ \\
     & Pretrained (weighted-avg) & $37.8^{\dagger}$ / $35.8^{\dagger}$ / $2.0^{\dagger}$ & $41.4^{\dagger}$ \\
     & Proposed & $\mathbf{37.4^{\dagger}}$ / $\mathbf{35.5^{\dagger}}$ / $\mathbf{1.9^{\dagger}}$ & $\mathbf{40.6^{\dagger}}$ \\
     \cmidrule(lr){1-4} 
     & None & $40.9$ / $37.0$ / $3.9$ & $49.9$ \\
     & Random  & $39.8$ / $37.2$ / $2.6^{\dagger}$ & $50.8$ \\
     Large & Pretrained (last-layer) & $35.8^{\dagger}$ / $33.2^{\dagger}$ / $2.6^{\dagger}$ & $45.3^{\dagger}$ \\
     & Pretrained (weighted-avg) & $35.3^{\dagger}$ / $32.8^{\dagger}$ / $2.5^{\dagger}$ & $42.8^{\dagger}$ \\
     & Proposed & $\mathbf{34.3^{\dagger}}$ / $\mathbf{32.2^{\dagger}}$ / $\mathbf{2.1^{\dagger}}$ & $\mathbf{42.6^{\dagger}}$ \\
    \bottomrule
  \end{tabular}
  
  \vspace{2pt}
  \footnotesize{$^{*}$p$<$.05,\; $^{\dagger}$p$<$.01: significantly lower than None by one-sided paired t-test}
\end{table}

\begin{table}[t]
  \caption{Ablation on diarization head using ADOS data. None means no diarization head is attached. Proposed uses pretrained (weighted average) with silence suppression.}
  \label{tab:dh-ablation-ados}
  \centering
  \begin{tabular}{l l c c}
    \toprule
    \textbf{Whisper} & 
    \textbf{Diarization Head} & \textbf{mtWER/WER/AER} & \textbf{DER} \\
    \cmidrule(lr){1-1} \cmidrule(lr){2-4}  
     & None & $30.0$ / $28.2$ / $1.8$ & $23.3$ \\
     & Random & $29.9$ / $28.1$ / $1.8$ & $25.5$ \\
     Small & Pretrained (last-layer) & $30.1$ / $28.6$ / $1.5$ & $22.8$ \\
     & Pretrained (weighted-avg) & $29.3^*$ / $28.3$ / $1.1^{\dagger}$ & $23.6$ \\
     & Proposed & $\mathbf{28.8^{\dagger}}$ / $\mathbf{27.8^*}$ / $\mathbf{1.0^{\dagger}}$ & $\mathbf{21.8^{\dagger}}$ \\
     \cmidrule(lr){1-4} 
     & None & $21.7$ / $20.3$ / $1.5$ & $19.3$ \\
     & Random & $21.3$ / $\mathbf{19.9}$ / $1.5$ & $19.1$ \\
     Large & Pretrained (last-layer) & $21.3$ / $20.1$ / $1.3$ & $\mathbf{18.2^{\dagger}}$ \\
     & Pretrained (weighted-avg) & $\mathbf{21.0^*}$ / $20.0$ / $\mathbf{1.1^*}$ & $18.5^*$ \\
     & Proposed & $21.7$ / $20.6$ / $\mathbf{1.1^*}$ & $18.4^*$ \\
    \bottomrule
  \end{tabular}
  
  \vspace{2pt}
  \footnotesize{$^{*}$p$<$.05,\; $^{\dagger}$p$<$.01: significantly lower than None by one-sided paired t-test}
\end{table}

Tables~\ref{tab:dh-ablation-play} and \ref{tab:dh-ablation-ados} report the effect of the diarization head on Playlogue and ADOS.
On Playlogue, jointly training a diarization head from scratch slightly improves MtWER, WER, and AER significantly relative to the ASR-only baseline, showing that adding the diarization objective does not harm recognition performance. However, DER is not significantly reduced, indicating that the non-pretrained diarization head provides limited diarization benefit. In contrast, the pretrained head improves all metrics, lowering mtWER and DER. With silence suppression, the improvements are largest, achieving the best mtWER, WER, and DER, and the lowest AER.

On ADOS, the diarization head without pretraining worsens DER. We found that when the diarization head is randomly initialized and trained on limited child–adult speech data, its optimization lags behind that of the ASR component, leading to incomplete convergence. As diarization head optimization is slower and incomplete without pretraining, noisy gradients from the randomly initialized head can lead to worse encoder speaker representations. In contrast, pretraining the diarization head yields a significant reduction in AER. Silence suppression further reduces the errors on Whisper-small models, while we show limited improvement for Whisper-large models, likely because the decoder-based diarization performance without silence suppression is already very strong (outperforming the diarization-first approach as shown in Table~\ref{tab:main-results}) and silence suppression may introduce additional errors.

Comparing the two pretraining strategies for the diarization head, pretraining with the weighted-average encoder representations consistently shows small improvements over the last-layer counterparts, while the overall performance remains comparable across the two datasets. Overall, these results demonstrate that joint ASR-diarization training, when combined with a pretrained diarization head, maintains the WER comparable to or better than the ASR-only baseline, while substantially reducing AER. Decoder-level silence suppression further improves the decoding robustness by producing cleaner speaker boundaries, especially under the noisier Playlogue dataset.

\subsection{Ablation on State-Machine-Based Forced Decoding}

\begin{table}[t]
\setlength{\tabcolsep}{3pt} % default is ~6pt
  \caption{Error Analysis with and without Forced Decoding (FD). P-FD and F-FD denote partial and full state-machine-based decoding, respectively. Failure rates (\%) are computed per single decoding run.}
  \label{tab:error-analysis}
  \centering
  \begin{tabular*}{1\linewidth}{l l c c c c c c}
    \toprule
    \multirow{2}{*}{\textbf{Whisper}} & \multirow{2}{*}{\textbf{Error}} & \multicolumn{3}{c}{\textbf{Playlogue}} & \multicolumn{3}{c}{\textbf{Ados}} \\ 
    & & \textbf{w/o FD} & \textbf{P-FD} &\textbf{F-FD} & \textbf{w/o FD} & \textbf{P-FD} & \textbf{F-FD} \\
    \cmidrule(l){1-2} \cmidrule(lr){3-5} \cmidrule(lr){6-8}
    \multirow{4.5}{*}{Small} & Miss Speaker & $0.1\%$ & $0.3\%$ & $0\%$ & $0.4\%$ & $1.0\%$ & $0\%$ \\
    & Miss Timestamp & $13.1\%$ & $0\%$ & $0\%$ & $16.0\%$ & $0.7\%$ & $0\%$ \\
    & Miss Both  & $4.2\%$ & $0\%$ & $0\%$ & $30.2\%$ & $0\%$ & $0\%$ \\
    \cmidrule(r){2-2} \cmidrule(lr){3-5} \cmidrule(lr){6-8}
    & Infinite Loop & $2.5\%$ & $1.2\%$ & $0.9\%$ & $12.3\%$ & $1.5\%$ & $1.0\%$ \\
    \midrule
    \multirow{4.5}{*}{Large} & Miss Speaker & $0.3\%$ & $0.3\%$ & $0\%$ & $1.3\%$ & $1.3\%$ & $0\%$ \\
    & Miss Timestamp & $1.0\%$ & $0.9\%$ & $0\%$ & $0\%$ & $0\%$ & $0\%$ \\
    & Miss Both  & $0\%$ & $0\%$ & $0\%$ & $0\%$ & $0\%$ & $0\%$ \\
    \cmidrule(r){2-2} \cmidrule(lr){3-5} \cmidrule(lr){6-8}
    & Infinite Loop & $1.4\%$ & $1.6\%$ & $1.7\%$ & $0.4\%$ & $0.4\%$ & $0.4\%$ \\
    \bottomrule
  \end{tabular*}
\end{table}

To better understand the effect of the proposed forced decoding mechanism, we analyze the major decoding failures observed in SOT-style generation: (1) missing only speaker tokens, (2) missing only timestamp tokens, (3) missing both speaker and timestamp tokens, and (4) infinite decoding loops. Table~\ref{tab:error-analysis} reports the percentage of decoding runs affected by each error on Playlogue and ADOS, comparing no forced decoding, partial forced decoding (P-FD; enforcing only the initial timestamp token), and the proposed full state-machine forced output decoding (F-FD). For all experiments in this ablation, we use models trained jointly with a diarization head and pretrained with a weighted average of Whisper encoders, and we do not apply silence suppression.

Without forced decoding, missing-token errors dominate for Whisper-small, with around 46.6\% and 17.4\% of decoding processes missing required speaker or timestamp tokens on the ADOS and Playlogue datasets, respectively. Importantly, these failures are largely driven by incorrect prediction of the very first token, which is expected to be a timestamp token by design. This issue is especially pronounced in the Whisper-small model, which often emits text tokens when it is uncertain about the initial timestamp, whereas Whisper-large does not exhibit first-token errors and adheres to the SOT format from the start. Empirically, once a correct timestamp token is generated, the remainder of the sequence is typically well-formed, suggesting that the first token serves as a critical anchor for subsequent structured generation.

Partial forced decoding (P-FD), which constrains only the first token to be a timestamp, already resolves the majority of these errors for Whisper-small. By enforcing a correct initial structural token, P-FD nearly substantially reduces missing timestamp or speaker token errors across both datasets, highlighting the importance of early decoding decisions in guiding the model toward consistent SOT formatting. However, P-FD does not completely eliminate the formatting errors later in the sequence, indicating that additional constraints are needed to fully ensure structural validity.

Full state-machine-based forced decoding (F-FD) further enforces all structural constraints and completely eliminates missing timestamp or speaker token errors, guaranteeing valid SOT outputs in every case. As a result, all utterances remain properly formatted, directly improving mtWER and timestamp-based alignment. The only remaining failure mode under forced decoding is the occasional occurrence of infinite loops, which persist at similar rates for both P-FD and F-FD.

Overall, these results demonstrate that a minimal forced decoding for the first token ensures that decoding begins with the correct format, while the proposed full forced decoding ensures complete structural validity. This reliability is crucial for downstream segmentation and role attribution, and it directly leads to more stable diarization-aware ASR performance.

\section{Applications}

\begin{table}[t]
\centering
\caption{Comparison between ground-truth (G.T.) and predicted (Pred) speech metrics on the Playlogue test dataset (N=34). PCC refers to the Pearson correlation coefficient.}
\begin{tabular}{lrrr}
\toprule
\textbf{Metric} & 
\textbf{G.T. Mean} & 
\textbf{Pred\ Mean} & 
\textbf{PCC} \\
\midrule
\multicolumn{4}{l}{\textit{Speech Quantity}} \\
Words per minute                & 42.64 & 41.44 & .975 \\
Utterances per minute           &  14.51 &  11.57 & .923 \\
\\[-0.9em]
\multicolumn{4}{l}{\textit{Utterance Length}} \\
Mean words per utterance        &  2.89 &  3.52 & .857 \\
Mean utterance duration (s)     &  0.80 &  1.15 & .531 \\
\\[-0.9em]
\multicolumn{4}{l}{\textit{Fluency}} \\
Speaking rate (words/s)         & 217.0 &184.11 & .645 \\
\bottomrule
\end{tabular}
\label{tab:metric_agreement_paylogue}
\end{table}

\begin{table}[t]
\centering
\caption{Comparison between ground-truth (G.T.) and predicted (Pred) speech metrics on the ADOS test dataset (N=84). PCC refers to the Pearson correlation coefficient.}
\begin{tabular}{lrrr}
\toprule
\textbf{Metric} & 
\textbf{G.T. Mean} & 
\textbf{Pred\ Mean} & 
\textbf{PCC} \\
\midrule
\multicolumn{4}{l}{\textit{Speech Quantity}} \\
Words per minute                & 48.68 & 50.63 & .971 \\
Utterances per minute           &  9.44 &  9.31 & .831 \\
\\[-0.9em]
\multicolumn{4}{l}{\textit{Utterance Length}} \\
Mean words per utterance        &  5.10 &  5.37 & .890 \\
Mean utterance duration (s)     &  1.84 &  1.95 & .704 \\
\\[-0.9em]
\multicolumn{4}{l}{\textit{Fluency}} \\
Speaking rate (words/s)         &163.58 &163.08 & .859 \\
\bottomrule
\end{tabular}
\label{tab:metric_agreement_ados}
\end{table}

\subsection{Automatically Derived Conversational Speech Metrics}

We analyze automatically predicted speech metrics that can be directly derived from the model-generated transcripts. For each child, we compute five speech measures grouped into three functional categories: (1) \emph{words per minute} and (2)
\emph{utterances per minute}, which together capture overall speech
\emph{speech quantity}; (3) \emph{mean duration per utterance} and (4) \emph{mean
words per utterance}, reflecting overall \emph{utterance length}; and (5)
\emph{speaking rate}, capturing speech \emph{fluency}. For the ADOS data, we merge the two sessions (\textit{Social Difficulties and Annoyance} and \textit{Emotional}) for the analysis. The test set consists of 84 children from MICH, all of whom completed both sessions. For the Playlogue dataset, we also evaluate on the test split at the child level, which includes 34 children.

Tables~\ref{tab:metric_agreement_paylogue} and \ref{tab:metric_agreement_ados} summarize the agreement between predicted
and ground-truth speech metrics for the Playlogue and ADOS datasets, respectively. For each measure, we report the mean value across children for ground-truth (G.T.) and predicted (Pred) sources, along with
the Pearson correlation (PCC) computed across children.
The predicted metrics closely approximate the ground-truth values across
all measures, with PCC ranging from $0.53$ to $0.98$ for the Playlogue dataset and $0.70$ to $0.97$ for the ADOS dataset. Strong agreements with the ground truth are observed in the words per minute and mean words per utterance metrics across both datasets, indicating that the model accurately captures the number of words produced by each child. Mean utterance duration and speaking rate metrics show lower correlations, likely because the errors are further compounded by diarization timestamp errors. Nevertheless, the overall results confirm that the system robustly recovers child spoken language characteristics, which can be helpful for downstream clinical analysis.

\subsection{Discussion on Clinical Applicability}
The proposed joint ASR and speaker-role diarization framework enables scalable analysis of child–adult interactions by directly producing structured, speaker-attributed transcripts with temporal boundaries, eliminating the need for manual segmentation. This substantially reduces the annotation burden and facilitates large-scale, longitudinal analyses of children’s expressive language and conversational behavior. In the context of child ASD research, prior studies \cite{kumar2016objective, eni2020estimating, mohanta2022analysis, assaf2025screening} have demonstrated that speech- and transcript-derived features are informative for modeling ASD-related behavioral variability. The proposed framework provides a practical foundation for extracting such features at scale. Future work can leverage the automatically generated speaker- and timestamp-tagged transcripts to develop clinically meaningful interactional and linguistic measures, and systematically connect these measures to clinician-annotated outcomes. While not a substitute for clinician-administered assessments, the proposed end-to-end joint modeling offers a promising pathway toward embedding automated speech and language analytics into clinical and research workflows involving child–adult interactions.

\section{Conclusion}
This paper has introduced a unified Whisper-based framework that jointly performs ASR and child–adult speaker-role diarization within a single end-to-end model. By combining serialized output training, timestamp prediction, a pretrained diarization head, silence-guided decoding, and a state-machine constraint, the system produces structurally reliable, speaker-attributed transcripts with accurate temporal boundaries. Experiments on Playlogue and ADOS-Mod3 show consistent improvements in mtWER and strong diarization accuracy compared with cascaded baselines. The model-derived conversational metrics closely align with those derived from human annotations. These results demonstrate the practicality and scalability of the proposed framework for automated analysis of child–adult interactions and related clinical research.

\section*{Acknowledgments}
This work was supported by \textsc{Simons Foundation (SFI-AR-HUMAN-00004115-03, 655054)}. We also thank Huang-Cheng Chou for insightful discussions and feedback.

\bibliographystyle{IEEEtran}
\bibliography{refs}

\end{document}